# Applying the Method of Critical Fluctuations on Human Electrocardiograms

**Yiannis Contoyiannis[(1)], Fotis Diakonos[(2)] and Myron Kampitakis[(3)]**

(1) Department of Electric-Electronics Engineering, West Attica University, 250 Thivon and P.Ralli, Aigaleo, Athens GR-12244, Greece (email: yiaconto@uniwa.gr)
(2) Department of Physics, University of Athens, GR-15874, Athens, Greece (email: fdiakono@phys.uoa.gr)
(3) Hellenic Electricity Distribution Network Operator SA, Network Major Installations Department, 72 Athinon Ave., N.Faliro GR-18547, Greece (email: m.kampitakis@deddie.gr)

**Abstract : In this work we apply the Method of Critical Fluctuations (MCF) on human Electrocardiogram (ECG) time-series. The method is able to reveal critical characteristics, in terms of physical behavior, in experimentally recorded signals. Using the concept of criticality as basic criterion for the characterization of the recorded ECG as that of a healthy person, we find a 100% verification of the characterization "Myocardial infarction". In contrary in the cases of the characterization "Healthy control" we find a 88% agreement. We also consider the autocorrelation function for the ECG time-series which obeys optimally the criteria of criticality and we observe the appearance of specific characteristic symmetries in the corresponding profile.**

The Electrocardiogram (ECG) is a basic resource of information about the heart operation. During the last years, many methods of "reading" ECG have been developed that can detect heart problems. Most of these methods are based on the differences between the structure of a recorded ECG and the prototype (control) structure of an ECG produced from healthy cardiac tissue. The classification of patients permits the correspondence of heart diseases in ECGs. Therefore, the revealing of the heart problem is possible through a simple reading of ECG. Today, the recognition of heart diseases can be accomplished through an Artificial Neural Network [1], usually with methods of image recognition. Nevertheless, there are cases where a simple "reading" of ECG is not enough to give a correct diagnosis about heart operation. The results of such diagnoses can become dangerous for patients. Therefore it is necessary to develop an analysis method which faces the ECG or the heartbeats as a time-series, allowing the application of a variety of methods. The most important methods of time-series analysis used are linear methods such as the FFT (Fast Fourier Transform) [2], Hurst-exponents analysis [3], Wavelet analysis [4] as well as non-linear methods, such as fractal analysis [5] and Hopf bifurcation [6]. In 2002 we had introduced an analysis method for time-series based in Physics of phase transition and critical phenomena in order to reveal a critical state. This method is named Method of Critical Fluctuations (MCF) [7] and has been applied in a variety of time-series produced in many systems such as Geophysics [8], simulated spin systems[7], Biological systems [9,10], Economical systems [11] and Electronics circuits [12].

In 2003 we applied the MCF in ECGs produced from frog heart [10]. The results of the MCF analysis are summed up in the phrase: When heart tissue is functioning properly (healthy state), then it is in a physical critical state. In Fig.1a segment of frog's ECG is presented.

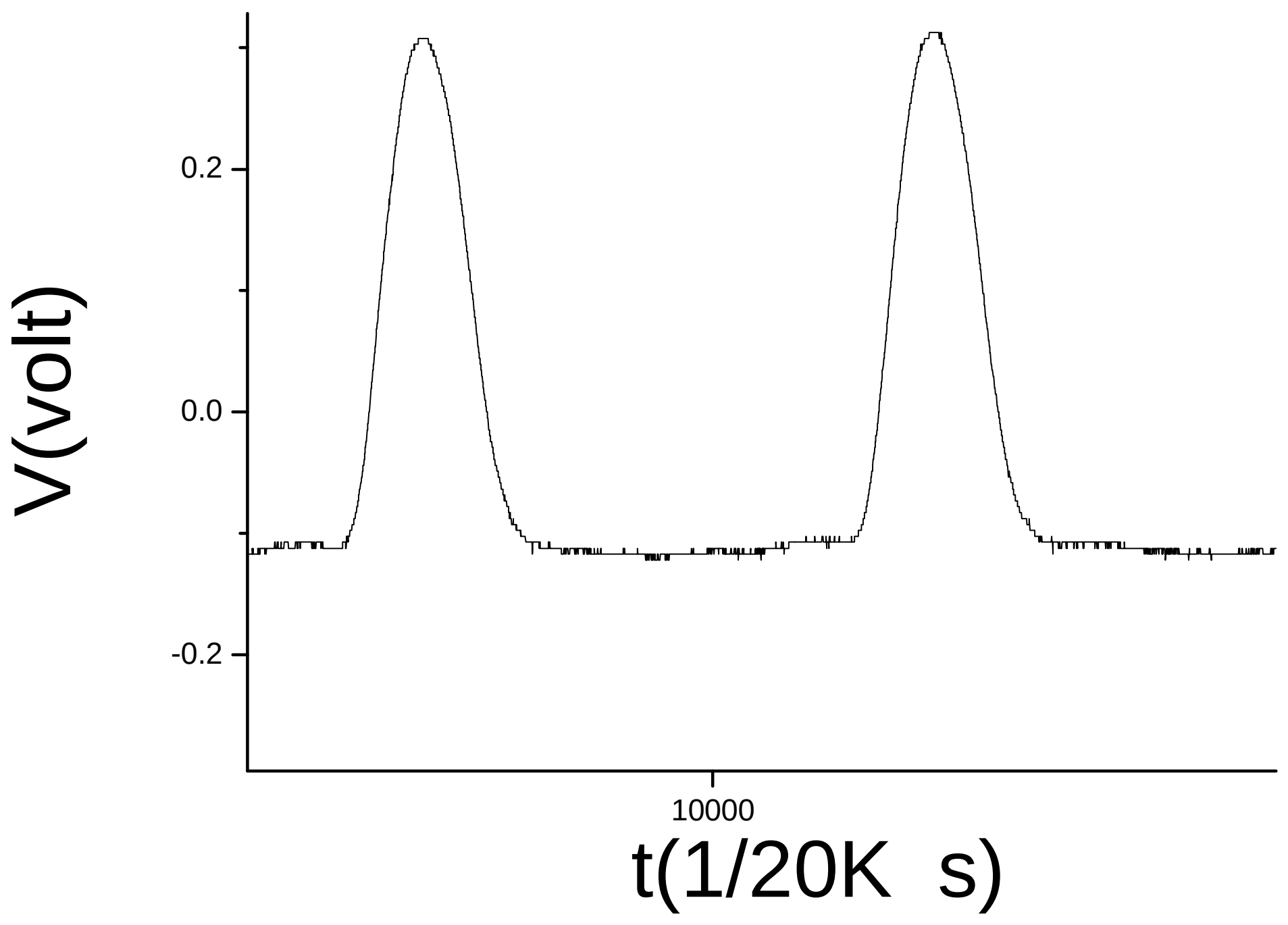

**Fig.1. A detail from frog's ECG. The fluctuations in relaxation period are shown**

The analysis is accomplished in a new time-series created by the relaxation phase where the fluctuations of ECG voltages appear. Criticality is present when the distribution of properly defined waiting times obeys power-law with exponent-values in the range [1,2] (see below). Basic property of criticality is that in critical state all the spatio-temporal scales of the system are present. The physical interpretation of critical state in terms of Physics in a healthy heart is that the tissue can respond to external stimuli in all time scales.

In the following, MCF is briefly presented. Details for this method are presented in [7,8,9]. Importantly, the exact dynamics at the critical point can be determined analytically for a large class of critical systems introducing the so-called critical map. This map can be approximated as an intermittent map:

$$\phi_{n+1} = \phi_n + u\phi_n + \varepsilon_n \qquad (1)$$

The shift parameter $\varepsilon_n$ introduces a non-universal stochastic noise which is necessary for the establishment of ergodicity [13]. Each physical system has its characteristic "noise", which is expressed through the shift parameter $\varepsilon_n$. Notice, for thermal systems the exponent z is connected with the isothermal critical exponent $\delta$ as $z = \delta + 1$. The crucial observation in this approach is the fact that the distribution $P(l)$ of the suitable defined laminar lengths $l$ (waiting times in laminar region) [7] of the above mentioned intermittent map of Eq. (11) in the limit $\varepsilon_n \to 0$ is given by the power law [14]

$$P(l) \sim l^{-p_l} \qquad (2)$$

where the exponent $p_l$ is connected with the exponent $z$ by $p_l = \frac{z}{z-1}$. Therefore the exponent $p_l$ is connected with the isothermal exponent $\delta$ by: $p_l = 1 + \frac{1}{\delta}$. The distribution of the laminar lengths of fluctuations is fitted by the function:

$$P(l) \sim l^{-p_2} e^{-p_3 l} \qquad (3)$$

We focus on the exponents $p_2$ and $p_3$. If the exponent $p_3$ is zero, then, the exponent $p_2$ is equal to the exponent $p_l$. The relation $p_l = \frac{z}{z-1}$ suggests that the exponent $p_l$ ( or $p_2$) should be greater than 1. On the other hand according to the theory of critical phenomena [15] the isothermal exponent $\delta$ is greater than 1. Therefore, as a result from $p_l = 1 + \frac{1}{\delta}$ we take $1 < p_l(p_2) < 2$. In conclusion, the critical profile of the temporal fluctuations is restored by the restrictions: $p_2 > 1$ and $p_3 \approx 0$. As the system removes from the critical state, the exponent $p_2$ decreases while simultaneously the exponent $p_3$ increases reinforcing, in this way, the exponential character of the laminar lengths distribution.

The frog's heart has three chambers and the structure of ECG is simple as it is shown in Fig.1. There is a relaxation period between two successive spikes. The MCF is applied for the segments of the relaxation period.

The human heart has four chambers and the ECG is more complex, as it is shown in Fig.2.

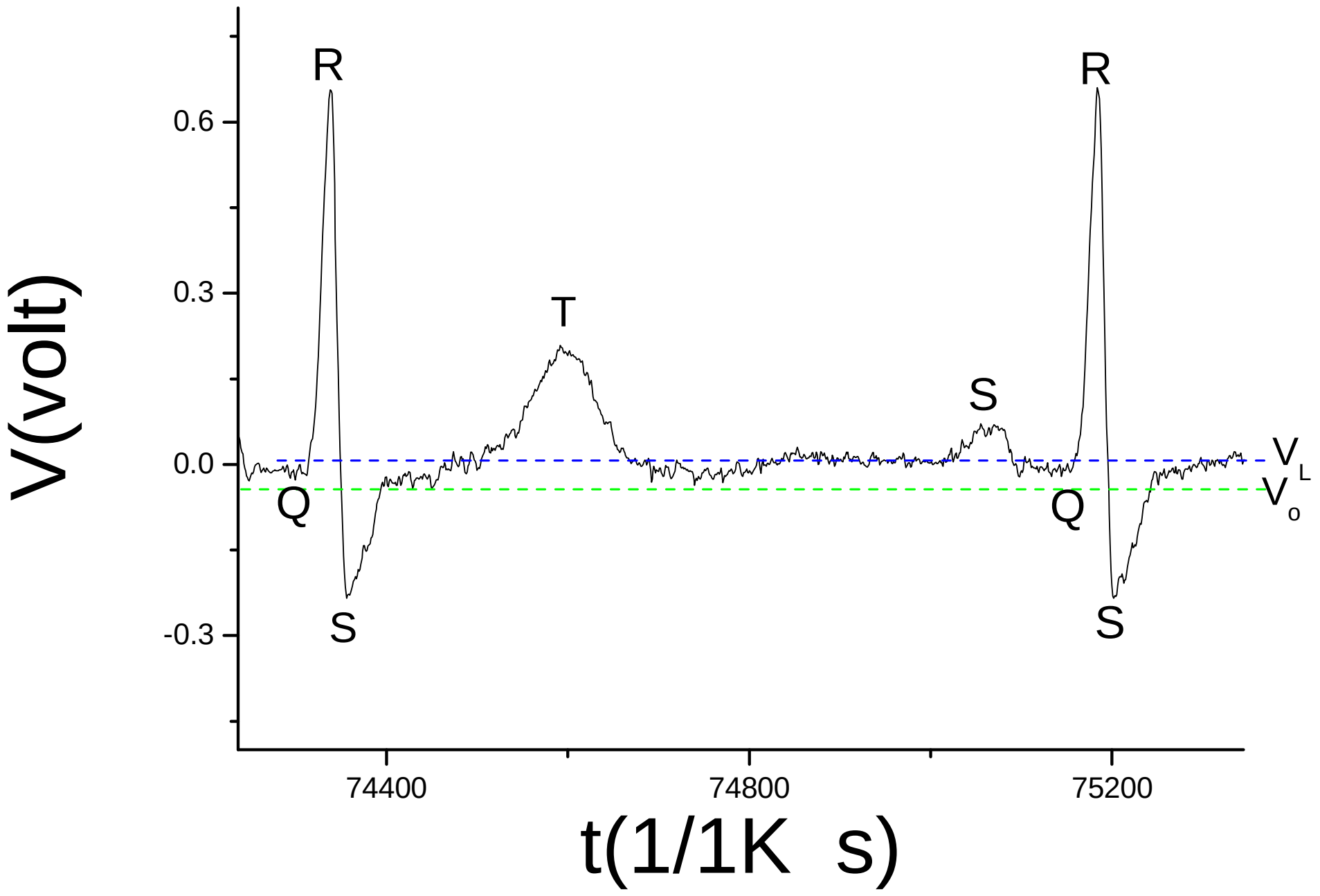

**Fig.2. In Human ECG the waves T,P and QRS complex wave are distinguished. Between the waves there are the intervals PR, ST, QT. The MCF analysis is applied where high frequency (grass) fluctuations appear.**

The critical dynamics in frog's ECG has been detected in the high frequency (grass) fluctuations. A further step was the extension of the MCF application in human heart. A first effort was made in our work [16], where we had presented the application of MCF in human ECG for the case of a healthy human heart and for the pathological case of a myocardial infarction. In the present work we analyzed more ECGs in order to obtain a statistical result. Moreover, a more detailed analysis is accomplished according to the more complex structure. The data (ECGs) were taken from the Physikalisch-Technische Bundesanstalt (PTB) Diagnostic ECG Database (Bousseljot et al., 1995; Kreiseler and Bousseljot, 1995; Goldberger et al., 2000) [17,18,19]. Each signal is digitized at samples with a frequency of 1000 samples per second (1 kHz). In Fig.3a segment of ECG from patient p.121 is shown. The selection of the segments of ECG is made with the criterion of stationarity , since a stationary structure of a time-series signal helps the MCF to converge.

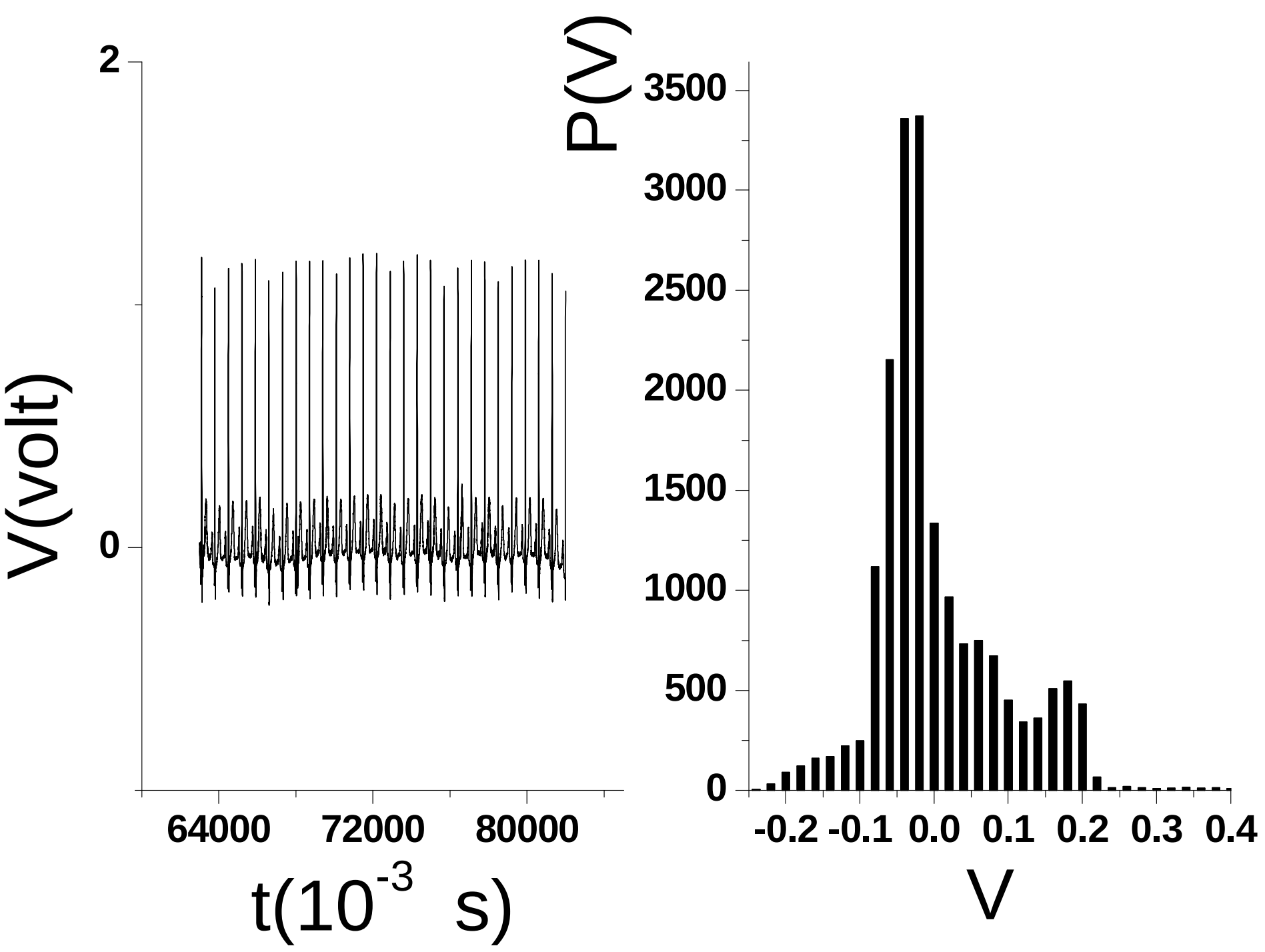

**Fig.3 (a). A segment of the ECG from p.121 is shown (sample length 19000 points). This segment has a good stationary structure. (b) The distribution of the values of ECG segment of Fig 3a.**

As it is shown in the distribution of Fig.3b, two lobes are present. The great lobe corresponds to the grass fluctuations around the isoelectric line of ECG and a small lobe corresponds mainly to the top of wave T. The second lobe does not appear in frog's ECG because the structure is simple and such a wave does not exist. If the segment length which obeys to stationarity is small, occurring in most cases, the tail of distribution (second lobe) has very small statistics and does not affect the results of MCF analysis. In the rare cases where the length of stationary segment is very large, we can produce two separated time-series, one for each lobe. The application of MCF on Human ECG follows the steps below:

- We localize the fixed point as the lower position where the grass fluctuations begin (green line in Fig.2).
- The zone between the fixed point level $V_0$ and the varied level $V_L$(blue line in Fig.2) define the laminar region. The condition $V_0$<$V_i$<$V_L$ defines the laminar lengths L (waiting times).
- We plot the distribution of laminar lengths L and we use the fitting function (3) in order to estimate the critical exponents $p_2, p_3$.
- We move the line $V = V_L$ to a new position and we repeat the second and third steps.

If we detect a zone $\Delta V_L$ where the distributions of laminar lengths corresponding to laminar region $[V_o, V_L]$ obey to the conditions of criticality (1<$p_2$<2, $p_3 \approx 0$ ), then it can be concluded that critical fluctuations in ECG appear. It is known that the larger the zone $\Delta V_L$, the more stable the critical state is [20].

In Fig.4 we present the MCF results from analysis on p.121.

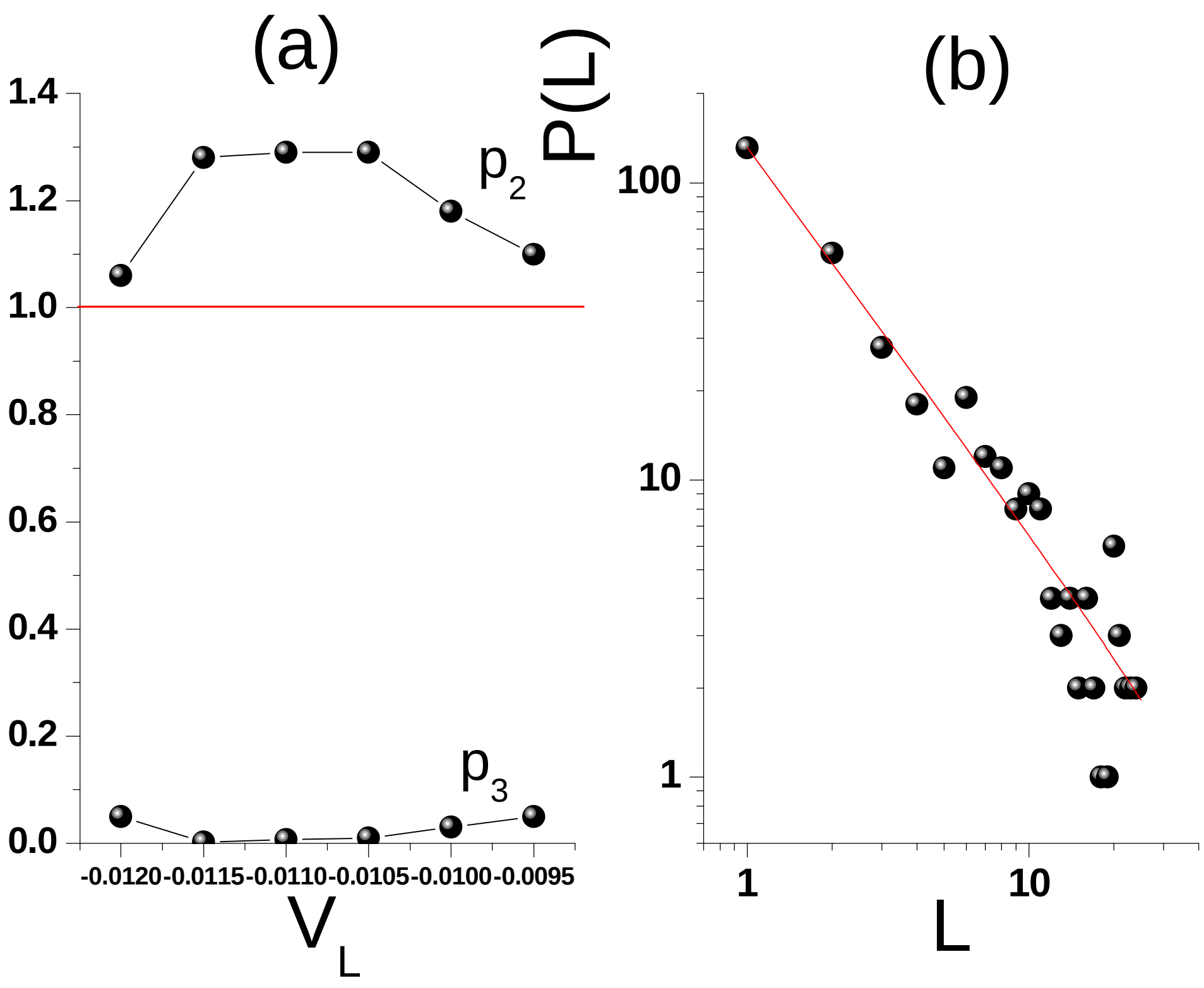

**Fig.4 (a) The $p_2, p_3$ values in the critical zone of laminar regions. The fixed point is $V_o$=-0.05 and the end of Laminar region $V_L$ is varied. (b) A representing example for $V_L$=-0.0115 where the laminar distribution is very close to a power law. The exponents are $p_2$=1.28, $p_3$=0.006 and $R^2$=0.990**

As we have mentioned above, the existence of criticality in ECG is a signature of the healthy state of cardiac tissue because it indicates that all the temporal scales are present in the heart response. The information from ECG waves is very suppressed due to small statistics in the tail of distribution of the ECG values. In the cases of time-series with a large length, if the stationarity criterion is satisfied, we can create a time-series including the second lobes, namely only the fluctuations of waves T,P (or only T in the case where height of P is small). Then we can apply the MCF for this time-series in order to deduce the information about the fluctuations.

We have analyzed 25 ECGs from the Diagnostic ECG Database [17,18,19] which have the indication “Healthy controls”. In 22 ECGs (88%) the MCF reveals critical fluctuations, in agreement with the characterization “Healthy control”, but in 3 cases (12%) our results disagree with the above characterization.

In the following, we proceed to the analysis of ECGs from “Myocardial infarction”. In Fig.5, such an example is presented.

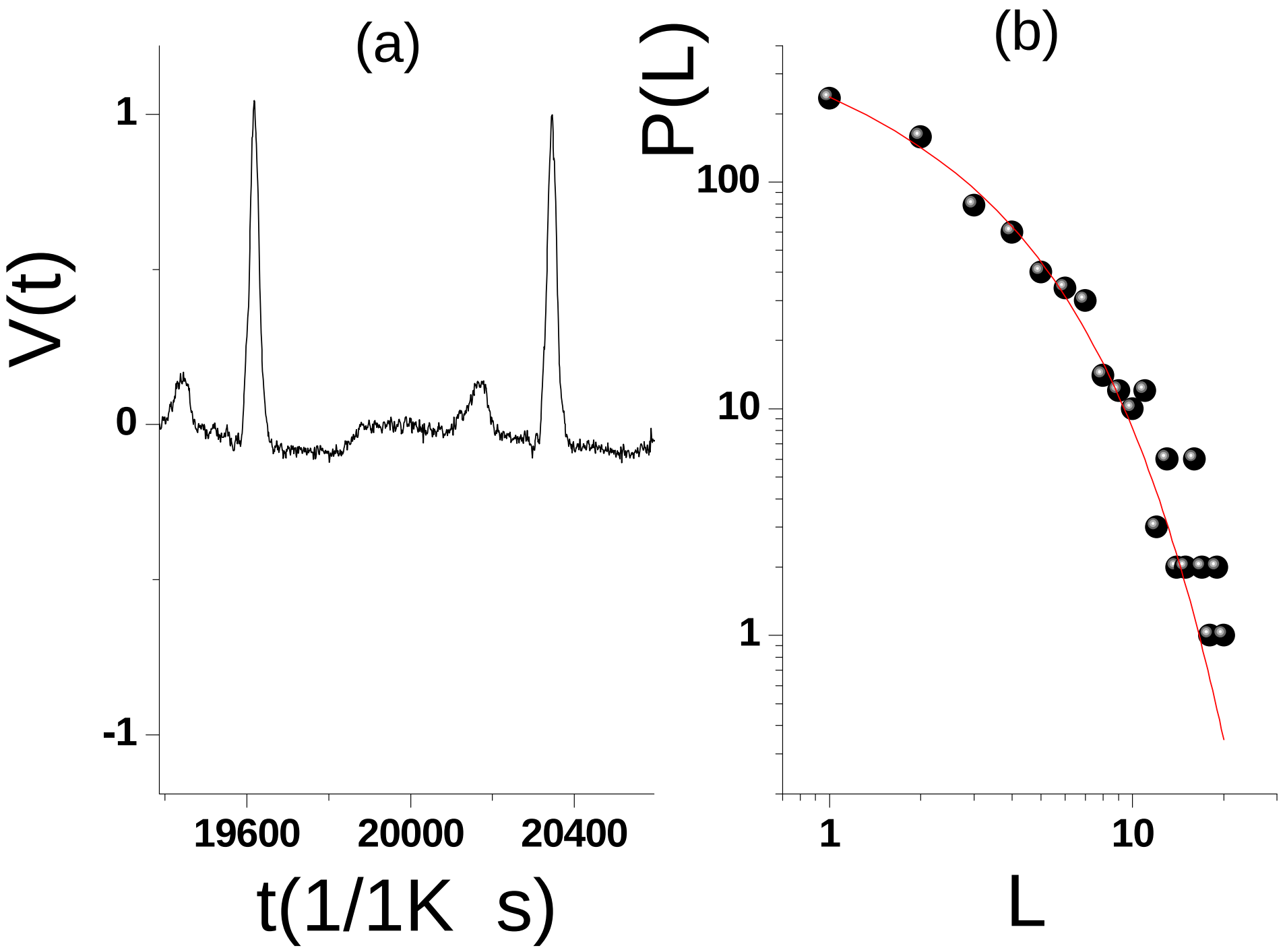

**Fig.5 (a) The ECG (a detail) for infarction case from patient p058. The distortion in the structure of ECG due to the infarction is obvious. (b) A typical example of laminar distribution from the application of MCF on the ECG of patient p058. It is obvious we are moving away from criticality. The estimated values are $p_2$=0.31, $p_3$=0.29, $R^2$=0.99.**

We have analyzed 10 ECGs from the Diagnostic ECG Database with Myocardial infarction. We conclude that all the cases are far from criticality, in absolute agreement with their characterization.

Next, we focus on the cases of the three patients with the characterization “Healthy control” where MCF results are out of criticality. To be more specific, the results of the analysis for the three patients (p165, p245, p242) exhibit values of exponents $p_2, p_3$ which are far from criticality, like the infarction case. In Fig.6 we present the results for p165.

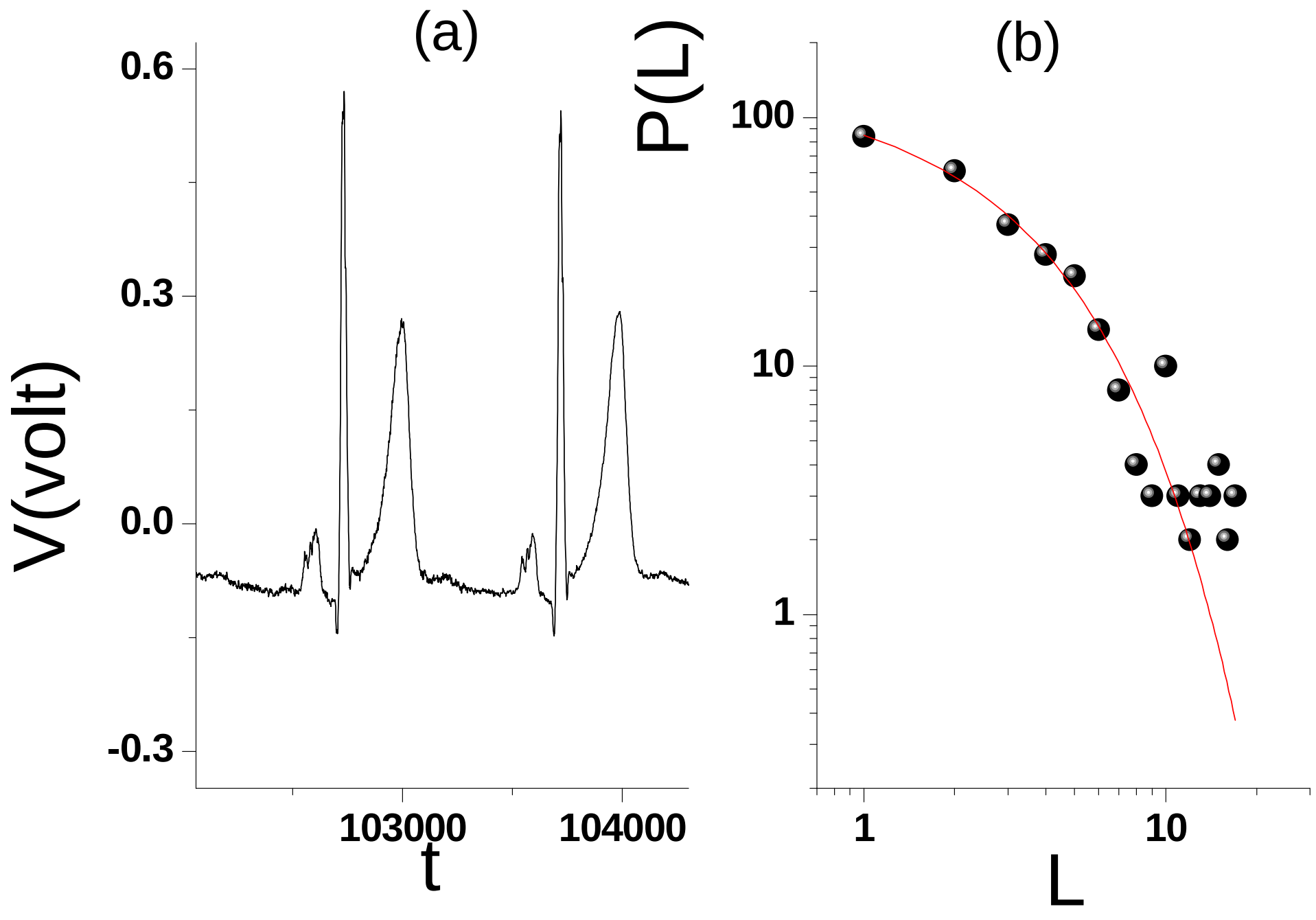

**Fig.6. (a) Detail from ECG p.165 characterized as control ECG from the Diagnostic ECG Database. The two basic waves T,P are present as in all the control signals. (b) A representative example of laminar length distribution from MCF analysis. The fitting line is very far from a power law. The estimated exponents are $p_2$=0.10, $p_3$=0.32. These results are similar with the results from analysis of infarction cases (Fig. 5b).**

The above behavior of one to ten control ECGs is really interesting and could be the object of further investigation.

There is a gradation in our results for criticality. This gradation is observed mainly in two characteristic parameters. The first one is the value of the $\boldsymbol{p_3}$ exponent which determines how close to the power law behavior is the distribution of laminar lengths. The closer to zero $\boldsymbol{p_3}$ is, the dynamics of fluctuations is closer to criticality and therefore ECG originates from healthier heart tissue. On the contrary, higher values of $\boldsymbol{p_3}$ mean that long time scales are cut in the heart tissue response. The second characteristic is the width of the zone $\Delta V_L$. As we mentioned above, as this width becomes greater, the stability of the healthy operation state of cardiac tissue is greater.

From all analyses in the present work, the case which best verified the above criteria about criticality was the p.121 presented in Fig.4. Inversely, p.121 could serve as a prototype definition of measures for the healthy behavior of cardiac tissue. For example, such a measure is the autocorrelation function which has a closerelationship with the critical phenomena [15]. In Fig.7 the autocorrelation function of p.121 ECG is presented.

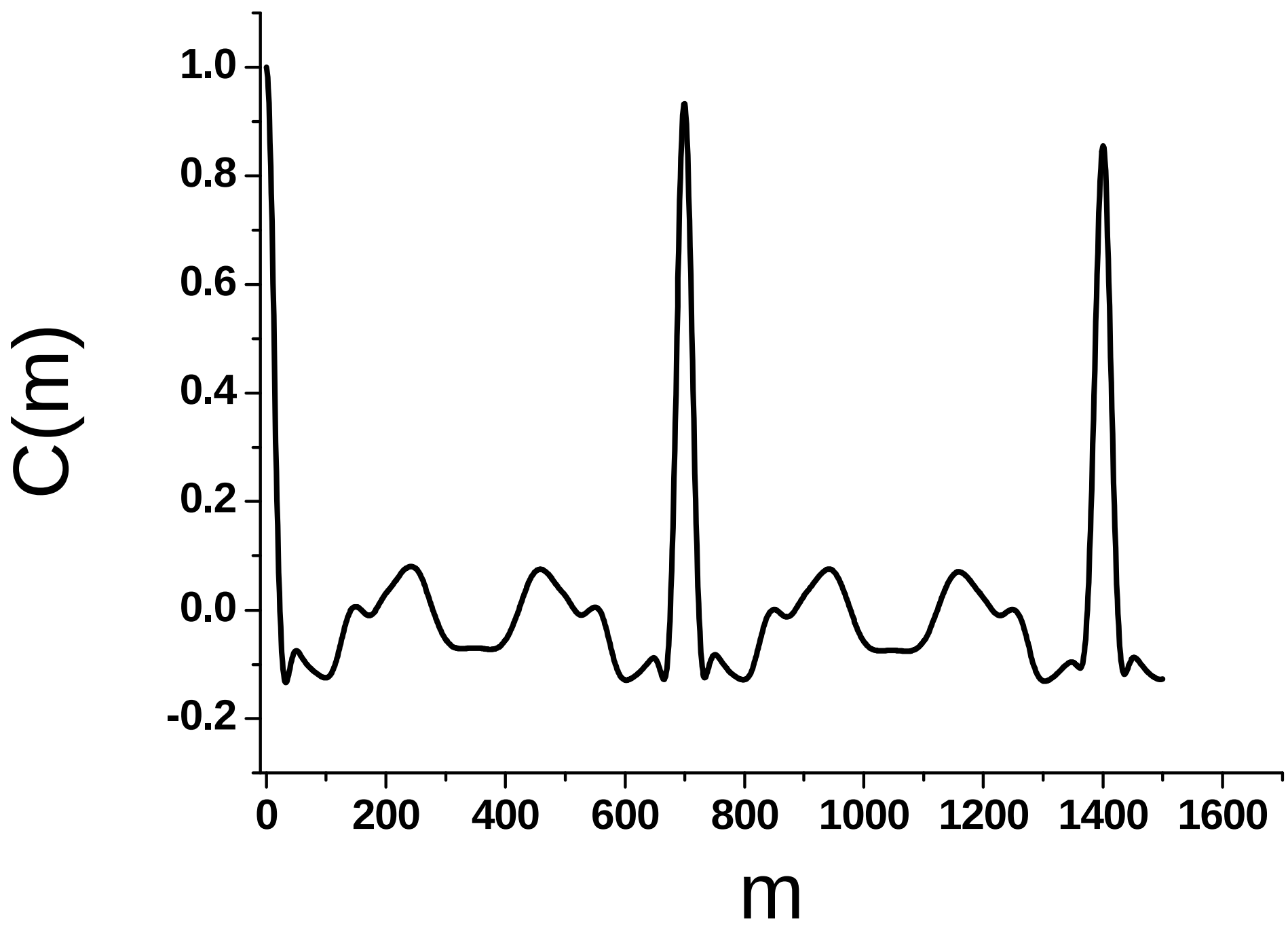

**Fig.7 The autocorrelation function of p.121. The symmetries which appear are due to the structure of the corresponding ECG.**

## Conclusions

In this work we presented the application of MCF analysis on Human ECG. The reference point of our work is that the existence of criticality in ECG fluctuations indicates the healthy operation of the cardiac tissue. The results of our analysis have shown that 88% of the samples of ECGs which are characterized as “Healthy control” exhibit fluctuations that obey to critical dynamics in terms of Physics. Nevertheless, there are 12% of the samples ECGs where our results do not agree with the characterization “Healthy control” because the dynamics of fluctuations is far from criticality. On the other hand, the results from 100% of the samples referring to “Myocardial infarction” cases showed that fluctuations are far away from criticality, namely there is 100% agreement with their characterization. Basic conclusion of our work is that in approximately 1 out of 10 ECGs which presented the typical characteristics of a healthy ECG, the diagnosis may not be accurate. Our method can reveal the healthy states as well as the pathological states. There are gradations between the ECGs that are characterized as healthy by our method. The criteria are the quality of power laws and the width of the zone $\Delta V_L$ of laminar regions. From the samples we analyzed , we detected one that best satisfied the criteria of criticality. For this sample we determined the autocorrelation function, which is considered to be a good measure for optimal heart functionality. Finally, it should be noted that this work interprets the heart operation in terms of Physical Critical Phenomena and is not involved with Medical Physiology. However, a link between the two descriptions is necessary and is the subject of further work.

## References

1. Himanshu Gothwal, Silky Kedawat, Rajesh Kumar "Cardiac arrhythmias detection in an ECG beat signal using fast fourier transform and artificial neural network", J.Biomedical Science and Engineering, 2011, 4, 289-296, doi:10.4236/jbise.2011.44039.
2. B.V.P Prasad & Velusamy Parthasarathy (2018)"Detection and classification of cardiovascular abnormalities using FFT based multi-objective genetic algorithm",Biotechnology & Biotechnological Equipment,32:1,183-193,DOI: 10.1080/13102818.2017.1389303.
3. Naoko Aoyagi*,* Ken Kiyono*,* Zbigniew R. Struzik*, and* Yoshiharu Yamamoto "Changes in the Hurst Exponent of Heart Rate Variability during Physical Activity", AIP Conference Proceedings 780, 599 (2005); https://doi.org/10.1063/1.2036824.
4. Alexander Z Tzabazis, Andreas Eisenried, David C Yeomans "Wavelet analysis of heart rate variability: Impact of wavelet selection**",** Biomedical Signal Processing and Control 40:220-225, February 2018*.* DOI: 10.1016/j.bspc.2017.09.027.
5. Jonathan Senand Darryl McGill "Fractal analysis of heart rate variability as a predictor of mFractal analysis of heart rate variability as a predictor of mortality: A systematic review and meta-analysis", Chaos 28, 072101 (2018); https://doi.org/10.1063/1.5038818.
6. Hrayr S Karagueuzian, Hayk Stepanyan and William J Mandel "Bifurcation theory and cardiac arrhythmias", Am J Cardiovasc Dis. 2013; 3(1): 1–16.
7. Y. Contoyiannis**,** F. Diakonos and A. Malakis**.** "Intermittent dynamics of critical fluctuations" Phys.Rev. Lett. 89, 35701(2002).
8. Y. Contoyiannis, P. Kapiris and K. Eftaxias**.** "Monitoringof a preseismic phase from its electromagnetic precursors". Phys. Rev. E 71, 1(2005).
9. Efstratios K. Kosmidis, Yiannis F. Contoyiannis , Costas Papatheodoropoulos ,Fotios K. Diakonos**"** Traits of criticality in membrane potential fluctuations of pyramidal neurons in the CA1 region of rat hippocampus ". European Journal of Neuroscience **,**https://doi.org/10.1111/ejn.14117 (2018).

10. Y. Contoyiannis, F. Diakonos, C. Papaefthimiou and G. Theophilidis. "Criticality in the relaxation phase of the spontaneous contracting atria isolated from the heart of the frog (Rana ridibunda)" Phys. Rev. Lett. 93, 098101 (2004)

11. A.Ozun, Y.Contoyiannis, F.Diakonos, L.Magafas and M.Hanias"Intermittency in Stock Market Dynamics".Journal of Trading. Summer 2014, Vol 9, No 3, pp26-33.

12. Stelios M. Potirakis , Yiannis Contoyiannis, Fotios K. Diakonos, and Michael P. Hanias"Intermittency-induced criticality in a resistor-inductor-diode circuit". Physical Review E 95, 042206 (2017).

13. Y.F. Contoyiannis and F.K. Diakonos . "Unimodal maps and order parameter fluctuations in the critical region" Phys. Rev. E 76,031138 (2007).

14. H. Schuster, " Deterministic Chaos" Weinheim, Germany: VCH, 1998

15. K. Huang "Statistical Mechanics" 2nd New York: Wiley 1987

16. Y. Contoyiannis, S.M. Potirakis, K. Eftaxias, "The Earth as a living planet: Human-type diseases in the earthquake preparation process" Nat. Hazards Earth Syst. Sci 13(2013) 125-139.

17. Bousseljot, R., Kreiseler, D., and Schnabel, A.: Nutzung der EKG-Signal datenbank CARDIODAT der PTB uber das Internet, Biomedical Engineering/Biomedizinische Technik, 40, 317–318, 1995.

18. Kreiseler, D. and Bousseljot, R.: Automatisierte EKG-Auswertung

Mit Hilfe der EKG-Signaldatenbank CARDIODAT der PTB, Biomedical Engineering/Biomedizinische Technik, 40, 319–320, 1995.

19. Goldberger, A. L., Amaral, L. A. N., Hausdorff, J. M., Ivanov, P. C., Peng, C.-K., and Stanley, H. E.: Fractal dynamics in physiology: Alterations with disease and aging, Proc. Natl. Aca. Sci., 99, 2466–2472, 2002.

20. Y.Contoyiannis, S.M.Potirakis, K.Eftaxias, M.Hayakawa, A.Schekotov "Intermittent criticality revealed in ULF magnetic fields prior to the 11 March 2011 Tohoku earthquake ( $M_w$=9)", Physica A 452(2016)19–28.